\documentclass{cupbook}
\usepackage{natbib}

\begin{document}

 \newcommand{\um}[1]{\"{#1}}

 \newcommand{\uck}[1]{\o}
 \renewcommand{\Im}{{\protect\rm Im}}
 \renewcommand{\Re}{{\protect\rm Re}}
 \newcommand{\ket}[1]{\mbox{$|#1\protect\rangle$}}
 \newcommand{\bra}[1]{\mbox{$\protect\langle#1|$}}
 \newcommand{\proj}[1]{\mbox{$\ket{#1}\bra{#1}$}}
 \newcommand{\expect}[1]{\mbox{$\protect\langle #1 \protect\rangle$}}
 \newcommand{\inner}[2]{\mbox{$\protect\langle #1 | #2
\protect\rangle$}}

\begin{center}
{\Large \bf Speakable and Unspeakable, Past and Future} \\
\vspace{0.2 in}
Aephraim M. Steinberg \\
Department of Physics\\
University of Toronto\\
Toronto, ONT M5S 1A7 \\
CANADA\\
(version of \today)
\end{center}
\vspace{0.5 in}

\section{Introduction}

A volume in honour of a visionary thinker such as John Archibald 
Wheeler is a rare license to exercise in the kind of speculation
and exploration for which Wheeler is famous, but which most of the
rest of us usually feel we had better keep to ourselves.  We have all -- 
even those of us who never had the fortune to work directly with 
him -- been inspired and motivated by Wheeler's creativity and 
open-mindedness.  For all of our apparent understanding of quantum
mechanics, our ability to calculate remarkable things using this 
theory, and the regularity with which experiment has borne out these 
predictions, at the turn of the 21st century it seems there are as 
many puzzles on the road to a true {\it understanding} of quantum 
theory as there were at the start of the previous century.  Then, at 
least, one could hope to be guided by the mysteries of unexplained 
experiment.  Now, by contrast, we may seem to have lost our way, as 
even though our experiments are all ``explained'' (in some narrow 
sense which can only be deemed satisfactory out of fear to leap beyond 
the comfortable realm of formalism), the theory itself is mysterious.  Further 
explorations, without the anchor of experiment, certainly run the 
risk of becoming mere flights of metaphysical fancy, giving rise to 
factions characterized less by intellectual rigor than by
fundamentalist zeal.  Yet it would be premature 
to give up the journey before at least trying to establish foothold 
on the terrain ahead.  Following Wheeler's example, we can invent new
experiments to help us speak about some of the unspeakable aspects of 
our theory, and to venture forward.

I have therefore decided to use this occasion to describe a number of 
loosely connected ideas we have been thinking about and experiments 
we have been working on in my group, which I believe relate to deep
 questions about how 
one should understand quantum mechanics.  In keeping with the best 
tradition, I provide no answers to these questions, but I hope 
that I can show how a variety of questions are related to one another, 
and related to experiments both {\it gedanken} and real.
Everything which 
follows takes place in the setting of standard quantum theory, and 
therefore even the most surprising predictions or observations I 
discuss are of course unambiguous, and implicit in every quantum 
textbook.  Why then are they surprising?  Clearly, we are not 
surprised only by results which contradict our theories; as is 
obvious when one discusses classical physics with students learning 
it for the first time, we are surprised by results which contradict 
what we {\it understand} of these theories.  Over and over again in 
the past decade or two, experiments in fields such as quantum optics 
have revealed phenomena which surprise even those of us who ought by 
now to know quantum theory reasonably well.  While many thinkers seem 
to consider such experiments mere parlor tricks, does not
the ability of these experiments to evoke continued surprise 
demonstrate that we still do not {\it understand} quantum theory the 
way we understand classical theory?  This simple observation is so 
clich\'ed as to bear repeating, for too many physicists have fallen 
prey to the reassuring but nihilistic thesis that since so many 
before us have failed, we would be wasting our time to seek any deeper 
understanding of quantum theory than is contained in our beautiful 
equations.  

\section{Past and future, particle and wave, locality and nonlocality}

\vspace{0.1in}
``Prediction is difficult, especially of the future.''
\vspace{0.1in}

This famous phrase is generally attributed to Yogi Berra, although 
among scientists one hears the credit given to Niels Bohr with some
frequency.  While the latter attribution has a certain comforting
believability to it, one wonders whether Bohr's theory would make
the past any more amenable to analysis than the future.
A moment's thought suffices to realize that as difficult as prediction
of the future may be, prediction of the past is not necessarily any
easier (even aside from the semantic issue, which leads us to adopt the 
term ``retrodiction'' for inferences about the past).  Neither is more
or less the domain of science, although physics has traditionally 
concentrated on prediction while fields such as archaeology and 
cosmology have dealt with retrodiction.  In classical mechanics, 
nevertheless, time-reversal symmetry guarantees that retrodiction 
is {\it precisely} the same task as prediction\footnote{For closed 
systems, at any rate-- the thermodynamic arrow of time breaks the 
symmetry in the case of open systems.}.  But in quantum mechanics as it 
is generally taught, despite the time-reversibility of the 
Schr\"odinger equation, retrodiction appears particularly mysterious.  
If I fire a photon towards a double-slit, quantum mechanics 
unambiguously tells me what the state of the photon is after passing
through the slits, although this state only gives probabilities for 
individual measurement outcomes.  But when I see the photon land at a 
particular point on the screen (see Fig. 1), 
what can I conclude about which slit 
it went through?  The usual approach to measurement, involving an 
uncontrollable, irreversible disturbance, effectively decouples the 
``collapsed'' state from what came before, except insofar as the 
probabilities for the measurement results are determined by the 
initial state.  This is quite different from the usual treatment of 
state preparation, which sets up a well-defined initial condition and 
allows unitary evolution to take over.

The orthodox view of quantum mechanics holds that what has been measured
can be known, and what has not is ``unspeakable.''  If a particle
is prepared in a certain wave packet, that function is to be considered
a complete description, and any additional questions about where the
particle ``is\footnote{Indeed, I once received an anonymous referee
report which read, in essence, ``This work is interesting, but 
I am unsure what the author means by the word `is'.''}" 
are deemed uncouth, at least until such a measurement is
made.  The absence of trajectories in quantum mechanics means that one
supposedly has no right to discuss where the particle ``was" prior to
that measurement.  Yet the fundamental laws of quantum mechanics are
as time-reversible as those of Newton, and one quite reasonably wonders
why it is any less valid to use a measurement to draw inferences about
a particle's history than to make predictions as to its future behaviour.
Such considerations led Yakir Aharonov and his coworkers to a formalism
of ``weak measurements" which allows one to discuss the state of evolving
quantum systems in a fundamentally time-symmetric way.  This chapter 
draws heavily on their ideas, whose main elements I will introduce 
below.  I will analyze how weak measurements can be applied to 
several experimentally interesting situations.  Consider, for one
example, the problem
of a tunneling particle.  What can we know about where a particle was
before it appeared on the far side of a forbidden barrier?  Is it
ever localized in the ``forbidden'' region?  Can we obtain more 
information about the particle's history from the state preparation, 
or from the observation that it was transmitted?

These new ideas about measurement naturally lead one to think about
epistemology.  Is the wave function the fullest description of what
we can know about a system?  Is there then a real sense in which a
particle may be in two places at the same time?
Can we sometimes have more information
than is encoded in a single wave function, by utilizing pre- and
post-selection simultaneously?  Or, on the contrary, is it impossible 
even
to know as much as a wave function, and are we limited to knowing the 
outcomes of the specific measurements we have performed?  Can we have
anything more than statistical knowledge about the outcomes of future
measurements?  Some experiments we plan to perform are designed to touch on
these issues.  In addition, they make one question whether even our
probabilistic description of reality is complete, or whether exotic
entities such as negative or complex probabilites may actually be
meaningful.

The explosive growth of the field of quantum information, with its
potential applications and headline-making buzzwords, has surprised
many by turning ``philosophical" research programmes into timely,
relevant, and some suspect even lucrative projects.  These questions about 
past and future are no exception.  Some of our recent
work has involved the development of
a quantum ``switch," in which a single photon may be transmitted
or not, depending on whether or not a single other photon is present.
The thorn is that it is impossible to know whether either photon
was ever present in the first place\ldots as in many quantum optics
experiments, the outcome depends on conditions which can only be
measured after the fact.  On this new work, I have no philosophical
conclusions to draw: only a cautionary tale about how tricky these
quantum conundra remain even for those building the experiments, and 
a hope that others will help us learn how to think 
about our own experiments in new ways.

To come full circle, our first planned application 
of this ``switch" is to carry out an experimental investigation of quantum
reality first proposed by Lucien Hardy, extending ideas due to Elitzur
and Vaidman.  This experiment allows one to demonstrate that what at
first glance appears to be perfectly airtight reasoning about the history
of particles once they have been detected can lead to a seeming
contradiction.  More recently, it has been recognized that this 
contradiction can be eliminated if one applies the formalism of weak 
measurements and accepts these ``exotic" probabilities as a correct 
description of reality.  We believe that most if not all of these
ideas are now accessible in the laboratory.

\section{Weak measurements}

The question of what measurement is is of course one of 
those which has haunted quantum theory from the start.  Why does one 
thing occur and not another (let alone more than one)?  When is a 
measurement?  How does this relate to the arrow of time?  By thinking
carefully about retrodiction as well as prediction, some of these 
issues can be, if not resolved, then perhaps at least brought into
starker relief.  Aharonov {\it et al.} have led the way in 
generalising concepts of measurement in this direction 
\citep{Aharonov=1990}, 
with their formalism of ``weak measurement.''
 In particular, weak measurements 
allow one to put past and future on an equal footing-- and, better 
yet, to do something which is commonplace to any experimentalist and 
yet seemingly at odds with the usual machinery of quantum theory: to 
use one's knowledge of the initial {\it and} the final conditions of a 
system together to draw conclusions about what came in between.

If the task of deducing what happened {\it before} a measurement was 
made based on the result of that single measurement seems to conflict
with the standard prescriptions of quantum theory, this is because 
the measurement is postulated to irrevocably change the state of the 
system.  But what is the origin of this disturbance?
Let us leave aside any considerations of ``collapse'' for the time 
being, and think only about the effect of an interaction between some 
system to be studied, and some other quantum mechanical system which 
will serve as a ``pointer,'' or measuring device.  Amplification of 
the state of this pointer to the macroscopic realm, so that a human 
observer might take note of it, can happen at some later stage if 
necessary; for our purposes, the important questions about 
measurement can all be treated simply by considering the effects of 
this quantum mechanical interaction.

In the standard approach due to von Neumann%
\citep{von_Neumann=1955,von_Neumann=1983}, 
a measurement of a system 
observable $A_{s}$ can be effected via an interaction Hamiltonian
\begin{equation}
{\cal H}  = g(t) A_{s}\cdot P_{p}\;,
\end{equation}
where the time-dependence $g(t)$ allows the measurement to take place 
during a finite interval of time, and where $P_{p}$ is the canonical 
momentum of the pointer.  Since the momentum is the generator of 
spatial translations, the effect of this interaction is to displace 
the pointer position by an amount proportional to the value of $A_{s}$.
In particular, for suitably normalized $g(t)$, the expectation value 
of the pointer position will change by an amount which is proportional 
to the expectation value of $A_{s}$, and thus serves as a record of 
this measured value.  Naturally, the requirement for a ``good'' 
measurement is that the pointer position be sufficiently well-defined 
that for different eigenvalues of $A_{s}$, the final state of the 
pointer is measurably different\footnote{Clearly, in the case of
an observable with a continuous spectrum at least, one must be
more cautious in defining precisely which eigenvalues ought to
be distinguishable.}.  In this case, the pointer and the 
system become {\it entangled}, and the irreversibility of the 
measurement can be seen as arising from the effective decoherence of 
the system wave function when one traces over the state of the pointer.

The back-action on the system can be seen in another way, which is 
that the above Hamiltonian exerts an uncertain force on the system,
to the extent that $P_{p}$ is uncertain.  If the pointer were in an 
eigenstate of momentum, then the measurement interaction would be an 
entirely predictable, unitary evolution of the system, ${\cal H} 
\propto A_{s}$; no irreversibility would thereby be introduced.  Of 
course, if the pointer momentum were perfectly well defined, the 
pointer position would be entirely uncertain, and it would be 
impossible to observe a translation of the pointer.  No measurement 
would have occurred.

Aharonov {\it et al.} argue that it is reasonable to consider an 
intermediate regime, where {\it some} information is captured during 
a measurement interaction, yet where the disturbance on the system is 
limited.  Although this is not the textbook model of a quantum 
measurement, it is in fact a good model of how countless experiments 
are actually performed.  Frequently, measurements on individual 
systems have such large uncertainties that only by averaging over 
thousands or millions of trials can statistical information be 
extracted.

The theoretical idea of a ``weak measurement'' is then to carry out a 
von Neumann interaction, but with an initial pointer state which is 
so delocalized in position that no single measurement can determine 
with certainty the value of $A_{s}$.  On the flip side, this pointer 
may have such small uncertainty in momentum that the back-action on 
the system can be made arbitrarily small.  It is in fact 
straightforward to verify that under these conditions, instead of
entangling the system and pointer according to 
\begin{equation}
\ket{\Psi}_{s}\phi_{p}(x) \rightarrow \sum_{i} c_{i} 
\ket{\psi_{i}}_{s}\phi_{p}(x-ga_{i})
\end{equation}
(where the $\psi_{i}$ and $c_{i}$ are the eigenkets of $A_{s}$ and
their corresponding amplitudes,  
and $ga_{i}$ is the shift in the pointer 
wavefunction
$\phi_{p}$ which corresponds to an eigenvalue $a_{i}$), the system and pointer remain to
lowest 
order unentangled:
\begin{equation}
\ket{\Psi}_{s}\phi_{p}(x) \rightarrow 
\ket{\Psi}_{s}\phi_{p}(x-g\expect{A_{s}}) \; .
\end{equation}
On average, the pointer is displaced by an amount related to the 
expectation value of $A_{s}$, but since this shift is too small to
significantly modify the pointer state, the system is unaffected.

Importantly, this means that the original evolution of the particle
may continue, and one may ask not only about the correlations between 
the pointer position and the initial state of the system, but equally 
well about correlations between the pointer position and the state the 
system is {\it later} observed to be in.  One may quite generally ask 
what will happen to the pointer on those occasions where the system 
was prepared in state $\ket{i}$ before the measurement interaction, 
and later measured to be in some final state $\ket{f}$.  Using 
standard quantum theory, Aharonov and coworkers showed that the 
mean shift of the pointer position for this subensemble corresponds to 
a ``weak value'' of $A_{s}$ given by
\begin{equation}
\label{weak}
\expect{A_{s}}_{\rm wk} = \frac{\bra{f}A_{s}\ket{i}}{\inner{f}{i}} \; .
\end{equation}
Clearly, for the trivial case $f=i$, this reduces to the usual 
expression for an expectation value.  But for the more general
case, it is heartening to note that the initial and final states have 
equal importance for the measured value of $A_{s}$; one can learn as 
much about a particle's state by observing its future as by knowing 
its past.

There are many other striking properties of weak measurements which 
suggest that they are a powerful tool for analyzing a broad variety of 
physical situations, and also that there may be some deep physical 
meaning to these quantities themselves.  I will not go over these in 
detail, but \citep{Reznik=1995} and \citep{Aharonov=Muga} 
provide a deep analysis.  In many ways,
these values can be seen as a natural application of Bayesian 
probability theory to quantum mechanics 
\citep{Steinberg=1995PRL,Steinberg=1995PRA}, satisfying many of the 
natural axioms of probability theory.  More important, they describe 
the outcomes of {\it any} measurements which can be described using 
the (modified) von Neumann formalism, and therefore show a clear 
connection to physical observables, not to mention a unifying 
framework within which a broad class of experiments may be treated.
At the same time, they display a number of troubling features.  
Notably, the measured weak value need not be consistent with any 
physically plausible values of $A_{s}$; it need not even fall within 
this operator's eigenvalue spectrum.  More shocking still, some 
positive-definite quantities such as energy (or even probability)
may be measured to be negative \citep{Aharonov=1993NKE}.  In fact, weak 
values are in general complex numbers rather than reals.  As explained 
in some of the above references, this is not an entirely untenable 
state of affairs, and the physical significance of the real and 
imaginary parts of the weak value may be clearly identified.  Roughly 
speaking, the real part indicates the size of the physical shift in 
pointer position, the measurement result one expected classically 
from such a device.  The imaginary part indicates how much the {\it 
momentum} of the pointer will change as an unintended consequence of
the measurement interaction, and consequently, how large the 
back-action of the measurement on the system.

One of the truly exciting features of weak measurements is that 
simultaneous weak measurements may be made on non-commuting 
observables, and do not render each other impossible, or even modify 
each other's results.  For instance, if a particle is prepared in a 
eigenstate of some operator $B$ with eigenvalue $b_{j}$, then a weak 
measurement of $B$ is guaranteed to yield the value $b_{j}$, 
regardless of the postselection.  Similarly, if it is {\it 
postselected} to have an eigenvalue $c_{j}$ of some operator $C$, then 
a weak measurement of $C$ is certain to yield $c_{j}$, regardless of 
the preparation.  If both $B$ and $C$ are measured weakly between the 
preparation and the postselection, both of these values will be 
observed  (albeit as {\it average} shifts of a very uncertain
pointer position)-- even if $B$ and $C$ do not commute.
For that matter, if $B+C$ is measured, the 
result will be $b_{j}+c_{j}$, something which makes intuitive
``classical'' sense, but which one could never hope for in the
context of strong quantum measurements.  Such properties clearly hold out the 
tantalizing possibility of making more of reality ``speakable'' (in 
John Bell's term \citep{Bell=1987}) than we are usually led to believe.  
When we think about a particular system which survived from state 
preparation through postselection, should we merely think of the 
initial state evolving in a unitary fashion until the postselection 
induced a collapse, or should we think about its properties as 
depending on both pre- and post-measurements?  While the orthodox view 
may be that if no measurement is performed between preparation and 
postselection, the question is meaningless, it is thought-provoking 
that {\it any} von Neumann-style interaction which takes place at 
intermediate times, provided that it is not so strong as to 
irreversibly modify the system dynamics, will produce an effect whose 
magnitude is defined by this new formalism.  Such observations led to 
a variety of speculations about the ``reality of the wave function''
\citep{Aharonov=1993PRL} and to a general formulation of quantum 
mechanics via ``two-time wave functions'' \citep{Reznik=1995}.

Recently, a connection has been drawn between weak measurements and 
more widespread techniques for dealing with the quantum evolution of 
open systems \citep{Wiseman=2002}, and this has proved useful for explaining 
the ``negative-time correlations'' in a cavity QED experiment 
\citep{Foster=2000}.  Specifically, an experiment in Luis Orozco's group
designed to observe the 
evolution of an electromagnetic field after the detection of one 
photon also found interesting dynamics in the evolution of the field 
{\it before} the detection of a photon.  
Howard Wiseman pointed out that when 
the photodetection event is treated as a postselection, an extension 
of weak measurement theory can be fruitfully applied to understand 
this negative-time evolution, which had not previously been fully 
explained.

\section{A quantum-mechanical shell game}
While it was recognized from the outset that weak measurements could
yield anomalously large values, and the first (intentional!) experimental
implementation of weak measurements was a linear-optics experiment to 
demonstrate how a spin measurement could yield an apparently 
nonsensical value\citep{Ritchie=1991}, it was pointed out 
\citep{Steinberg=1995PRA} that there is a striking mathematical 
relationship between weak measurements and classical probability 
theory.  In fact, the result of Eq. \ref{weak} can be obtained quite 
generally by summing over the ``conditional probabilities'' for each 
of the eigenstates of the operator, 
\begin{equation}
\expect{A}_{\rm wk} = \sum_{j} a_{j} P(j | i,f) \; ,
\end{equation}
where the probability of being in 
an eigenstate $\ket{\psi_{j}}$ is defined as the expectation value of 
the projector $\proj{\psi_{j}}$;  the ``conditional probability'' 
is the natural generalisation based on the weak-measurement prediction 
for the shift experienced by a pointer which couples to this 
projection operator, conditioned on the appropriate post-selection:
\begin{equation}
P_{\rm wk}(j | i,f) = 
\frac{\inner{f}{\psi_{j}}\inner{\psi_{j}}{i}}{\inner{f}{i}} \; .
\end{equation}
Of course, these conditional 
probabilities sometimes prove to have values greater than 1, less than 
0, or even with imaginary components.  It is on the one hand unclear 
if it is meaningful in any real sense to interpret these as 
probabilities, while on the other hand the weak-value expressions for 
probability are defined in clear analogy to classical probabilities, 
and satisfy the same axioms.  Furthermore, the experiments which are 
predicted to yield negative or complex ``probabilities'' are 
designed in precisely the fashion one would choose classically to
measure the conditional probabilities, and they would correctly measure
these probabilities when used in the classical regime; is this not
the operational prescription for developing the quantum mechanical 
formalism for a given observable?

I do not possess the hubris to attempt
 to pronounce a final verdict on how seriously one 
should take these probabilities, or on whether one would be better to
avoid such a loaded term at all.  Nevertheless, the expressions 
derived in this fashion have clear physical significance for a 
wide-ranging class of experiments.  Suffice it to note that there are
a number of other contexts (such as ``rescuing'' locality, in the
context of Bell's theorem) in which other authors have suggested
taking seriously the concept of negative probability in quantum
mechanics \citep{Muckenheim=1983,Feynman=1987,Pitowski,Scully=1994}, 
not to mention the negative 
quasiprobabilities which are familiar in the context of the Wigner
function and other phase-space distributions \citep{Wigner=1932,%
Liebfried=1996}.

Let us for now accept this terminology of probabilities, with all its
caveats, and examine some striking examples of what weak-measurement
theory predicts.  In 1991, Aharonov and Vaidman applied the formalism 
to the following toy problem \citep{Aharonov=1991}.  Consider a particle 
which can be in any of three boxes, which we will denote as three 
orthogonal states $\ket{A}$, $\ket{B}$, and $\ket{C}$.  Let us 
prepare the particle in an initial state 
\begin{equation}
\ket{i} = \frac{\ket{A}+\ket{B}+\ket{C}}{\sqrt{3}} \; ,
\end{equation}
i.e., a symmetric equal superposition of being in each of the three 
boxes.  Suppose that some time later we choose another basis, and 
measure whether or not the particle is in the final state
\begin{equation}
\ket{f} = \frac{\ket{A}+\ket{B}-\ket{C}}{\sqrt{3}} \; ,
\end{equation}
where the sign in front of box $C$ has been changed.  Note that there
is some probability for this postselection to succeed, without any 
need for the particle to change its state between the measurements;
$|\inner{f}{i}|^{2}= 1/9$.  

Obviously, the question of interest is how we should describe the 
state of the particle between the state-preparation and a successful 
post-selection.  Should we evolve $\ket{i}$ forward in time under the 
free Hamiltonian, the particle remaining symmetrically distributed 
among the three boxes, until the final measurement disturbs its 
phase?  Or should we instead evolve $\ket{f}$ backwards in time?  
Clearly, orthodox quantum mechanics says there is no meaning to the
question of {\it at what time} $C$ stopped being in phase with $A$ 
or $B$, and began being out of phase with them; Bohr would tell us 
that the value of this phase during a period when nothing in the 
apparatus is sensitive to it is meaningless.  Similarly, we cannot 
ask which of the three boxes the particle was in before it was 
detected in $\ket{f}$, although it seems quite natural to suppose it 
had equal probabilities to be found in any of them.  

One can conceive of measuring such probabilities, by using a large 
ensemble of particles.  For instance, a test charge held near box A 
may experience a slight momentum shift if and only if the particle is 
in box A.  If this shift is arranged to be far smaller than the 
uncertainty in the test charge's momentum, then it may be possible to 
carry out such measurements without any appreciable effect on the 
evolution of the particle.  If no postselection is performed, the 
magnitude of this shift will be proportional to the probability that 
the particle was indeed in $A$, i.e., the expectation value of the 
projection operator $\proj{A}$.  For the state $\ket{i}$, for 
instance, this probability is one third: the impulse imparted to the 
test charge after $N$ particles go through the boxes will be 
precisely what one would expect if $N/3$ had been in box A\ldots or,
equivalently, if one third of {\it each} of the $N$ particles had been 
in box $A$.

What if the momentum shift on the test charge is recorded (including 
its large uncertainty) each time a particle passes, but is discarded 
unless the postselection fails?  Then the sum of the momentum shifts 
for all the test charges which interacted with particles eventually 
detected in $\ket{f}$ will describe the {\it conditional} probability 
that those particles had been in box $A$:
\begin{eqnarray}
P_{\rm wk}(A |i,f) & = & \frac{\bra{f}{\rm 
Proj(A)}\ket{i}}{\inner{f}{i}} \nonumber \\
& = & \frac{\inner{f}{A}\;\inner{A}{i}}{\inner{f}{i}} \; .
\end{eqnarray}
It is easy to verify that this probability is unity.  The postselected 
test charges will display precisely the same mean momentum shift as 
they would for a particle prepared with 100\% certainty in box $A$.
Similarly, the weak (or conditional) probability for the particle to 
be in box $B$ is 100\%.  And the axioms of probability?  Must not the 
probabilities of all the exclusive possibilities add up to $1$?  
Indeed-- it is equally easy to verify that $P_{\rm wk}(C|i,f)$, the
conditional probability for a particle to have been in box $C$ between 
its preparation in $\ket{i}$ and its detection in $\ket{f}$, is $-1$.
Meaningless?  Not at all.  If the mean momentum shift of test charges 
which interact with particles eventually detected in state $\ket{f}$ 
is measured, it will be found to have the ``wrong'' sign-- that is, 
if the particle and the test particle have charges of like signs and 
ought to repel each other, the test charge will be found to have a 
mean momentum {\it towards} box $C$.  Perhaps it is risky to interpret 
this by saying the particle truly had a negative probability to be in 
that box-- yet physically, its effect was equal and opposite to the 
effect of a particle in box $C$.

Perhaps more striking yet is the observation that the particle was 
{\it ``definitely''} in box $A$, but also in box $B$.  We are quite 
accustomed to saying that a particle must go through ``both slits at 
once'' in Young's interferometer, but how many of us truly mean it?  
The wave function, of course, traverses both slits, but we know full 
well that to talk of ``the'' position of the particle, we must 
introduce some position-measurement, in which case the particle will 
be observed at one slit or the other.  Weak measurements show us that 
this is not necessarily always the case, so long as no ``collapse'' 
(or decoherence, more precisely) is introduced during the 
measurement.  Aharonov {\it et al.} have used these features of the 
theory to argue in favour of the ontological ``reality of the wave 
function'' \citep{Aharonov=1993PRL}, while these arguments have 
incited a great deal of controversy \citep{Unruh=1994}.  More recently,
Aharonov and Vaidman have tried to respond to some objections to 
their shell-game paper by introducing a {\it strong} measurement-- 
they show that if this particle is a ``shutter,'' then a photon 
heading towards either box $A$ or box $B$, or indeed any superposition 
of the two, is guaranteed to be intercepted by the shutter (in cases 
where the shutter is postselected to be in $\ket{f}$, as 
always)\citep{AV=slits}.  This suggests that the nonlocality of quantum 
mechanics may be even deeper than usually recognized, in that a given 
particle could actually have measurable effects in two places at the 
same time.

We are currently setting up an experiment, shown schematically in 
Figure 2, designed to test some of the features of this quantum 
conundrum.  Photons are prepared in a symmetric superposition of the 
three ``boxes'' $A$, $B$, and $C$, by the use of beam-splitters; each 
box is in fact one path in an interferometer.  By carefully adjusting 
the relative phases of the paths (specifically, by introducing an 
extra $\pi$ phase shift along path $C$ before symmetrically 
recombining the three beams at another beam splitter), it is possible 
to project out light in the state $[\ket{A}+\ket{B}-\ket{C}]/\sqrt{3}$.
Several varieties of weak measurement may be performed.  In 
particular, a small piece of glass can introduce a spatial shift in 
one of the three beams, smaller than the width of the beam (i.e., the 
uncertainty in the photon's transverse position).  Alternatively, a 
waveplate can rotate the polarisation of one of the paths by a small 
angle.  It is an optics problem left for the reader to show that the 
deviations to be expected are precisely those predicted by weak 
measurement theory: if beam $A$ or $B$ is displaced by $\delta x$, 
then the output will be displaced by $\delta x$\ldots on the other 
hand, if beam $C$ is displaced by the same amount, the displacement at 
the output will be $-\delta x$.  (In the optics context, it is not 
difficult to understand this as an interference effect related to 
the $\pi$ phase shift introduced in arm $C$.)  We plan not only to 
confirm the weak-measurement predictions, but also to study the {\it 
correlations} between the different probabilities.  In particular, we 
are interested in the question of nonlocality.  If we can say with 
certainty that the particle was in $A$ and that it was in $B$, can we 
also say that it was simultaneously in $A$ and $B$?  This may seem 
obvious, but again, with weak measurements one must be careful.

In their paper on ``How One Shutter Can Close $N$ Slits'' 
\citep{AV=slits}, Aharonov and Vaidman note that a {\it pair} of test
particles, one heading to shutter position $A$ and the other to 
shutter position $B$, could {\it not} both be reflected by a single 
shutter (although they make interesting observations about the case of 
multiple slits, multiple shutters, and multiple incident particles).  
In essence, the reflection of a particle heading towards $A$ is a 
strong measurement, and prevents the slit from stopping a second 
particle heading towards $B$.  However, one can put this even more 
succinctly if one accepts the definition
\begin{equation}
P(A \& B) = \expect{\rm Proj(A) \cdot Proj(B)} \; .
\end{equation}
Although this definition has certain pathologies associated with it 
\citep{Steinberg=1995PRA} (notably, this product of two projectors need 
not be a Hermitian operator, and therefore could yield complex ``joint 
probabilities'' even in non-post-selected systems), it seems the most 
natural way of describing joint probabilities.  It generalizes easily 
to the case of weak (conditional) measurements.  However, if $A$ and 
$B$ are orthogonal, as in the present case, then the product of their 
projectors
\begin{eqnarray}
{\rm Proj(A)\cdot Proj(B)}  =  \ket{A}&\inner{A}{B}&\bra{B} \nonumber \\
 =  \ket{A} & 0 & \bra{B} = 0 \; .
\end{eqnarray}
Under no circumstances is there a nonzero joint probability, 
conditional or otherwise, to be in box $A$ {\it and} to be in box 
$B$.  As discussed in \citep{Aharonov=2002}, weak measurements do not 
allow one to conclude that because $P(A)=P(B)=1$, then $P(A\& B)$
must also be $1$; this is because the probabilities themselves are 
not bounded by $0$ and $1$.  The probability of ``A and B'' may 
vanish, in spite of the certainty of $A$ and $B$ individually,
for the probability of ``A and not B'' is one.  If this seems 
strange, given that the probability of ``not B'' is zero, no worries: 
for the probability of ``not A and not B'' is negative 1.  This odd
state of affairs is summarized in table 1.

\section{Tunneling}

Another problem where nonlocality has been a topic of discussion in 
recent years is that of tunneling through a barrier.  It has been 
well-known since early in the century 
\citep{Wigner=1955,MacColl=1932,Hauge=1989,Buttiker=1982} that the group 
delay (stationary phase time) for a wave packet incident on an opaque 
barrier of thickness $d$ to appear on the far side saturates to a 
finite value as $d$ tends to infinity.  For large enough $d$, this 
implies superluminal propagation speeds for the peak of the wave 
packet, which naturally provoked much skepticism.  A number of 
experiments, including one I performed along with Paul Kwiat 
in Ray Chiao's group at Berkeley \citep{Steinberg=1993PRL}, 
demonstrated that this prediction is indeed correct 
\citep{Enders=1993,Spielmann=1994}, although no violation of causality is 
implied\citep{Chiao=1997}.  Due to the difficulty of timing the arrival of 
matter particles through any reasonable tunnel barrier, and the 
problems of reaching the relativistic regime with massive particles, 
these experiments were carried out with photons.  We are now building 
at Toronto a series of experiments designed to observe the tunneling 
of laser-cooled atoms through micron-scale barriers formed by focussed 
beams of light\citep{Steinberg=AnnPhys,Steinberg=1998SLM}.  Although the experiments are 
complex, this should open up a broad new vista of phenomena to 
study.  In particular, it becomes possible to {\it probe} the 
particles while they are traversing the ``forbidden'' region, and also 
to study the effects of decoherence on the tunneling process%
\citep{Steinberg=JKPS}.

\begin{figure}
\begin{center}
\begin{tabular}{c||c|c||c}
 {\it Probabilities}   & A & not A & A or not A\\
    \hline \hline
 B & 0 & 1 & 1 \\
 \hline
 not B & 1 & -1 & 0 \\
 \hline \hline
 B or not B & 1 &0 & \\
 \end{tabular}
 \end{center}
\caption{This table summarizes the probabilities and joint 
probabilities of finding the particle in or out of box $A$ and in or 
out of box $B$, demonstrating how a negative probability in one column 
can allow the joint probability of two ``certain'' events to vanish.}
\end{figure}

While it is certainly strange that a wave packet peak should arrive in 
less time than if the original peak had travelled at the speed of 
light, it was pointed out comparitively early in the (latest bout of 
the) tunneling time controversy that no physical law guarantees any 
direct causal connection (let alone identity) between an incoming peak 
and an outgoing peak \citep{Buttiker=1982}.  We generally interpret these 
effects as remarkable but entirely causal ``pulse reshaping'' 
phenomena, in which the leading edge of a pulse is preferentially 
transmitted, while its trailing edge is preferentially reflected, thus 
biasing the peak towards earlier times.  Similar effects had been 
observed in the 1980s in the context of propagation through absorbing 
media \citep{Garrett=1970,Chu=1982}, and much excitement has 
recently been created by the analogous observation of 
faster-than-light propagation in transparent (but active) media
\citep{Steinberg=1994FTL,Wang=2000,Steinberg=2000}.  A review of 
superluminality and causality in optics is given in \citep{Chiao=1997}.

These counter-intuitive effects occur only when the tunneling 
probability is relatively small.  In other words, like many 
weak-measurement paradoxes, the anomalies are dependent on the success 
of a postselection which occurs only rarely.  If one tracks the centre 
of mass of a wave packet incident on a tunneling barrier, it never 
moves faster than light-- only when one projects out the transmitted 
portion alone does the peak abruptly appear to have travelled 
superluminally.  In this sense, one may well argue that the 
superluminality is not a function of propagation through the tunnel 
barrier, but only of this mysterious ``collapse'' event whereby a 
particle previously spread out across two peaks may choose to localize 
itself on one.  Nevertheless, it seems reasonable to apply the 
formalism of weak measurements to the tunneling problem, in order to 
see whether this can shed light on the counterintuitive aspects of the 
situation.  For instance, can one verify that the tunneling particles 
originated predominantly near the peak of the wave packet?  Can one,
alternatively, determine the length of time a particle spends (on 
average) under the barrier?  This ``sojourn'' or ``dwell'' time is a 
quantity which had been of much interest to the condensed-matter 
community, as it would allow one to describe the importance of 
interactions between a tunneling particle and the surrounding 
environment, and the validity of approximations such as adiabatic 
following.  Even those who were not troubled by the superluminal {\it 
peak delay} presumed that the physical time spent in a given region 
of space would have to be greater than or equal to $d/c$; a number of 
models of the interaction between a tunneling particle and the 
environment were used to support this conjecture and yield 
``interaction times'' for the tunneling problem\citep{Buttiker=1985}.

In \citep{Steinberg=1995PRL,Steinberg=1995PRA}, I applied the ideas of 
weak measurement to this question, and was surprised.  On the one 
hand, no weak measurement would show the supposed ``bias'' towards 
the leading edge of the incident wave packet.  Furthermore, one could 
rewrite the tunneling ``interaction time'' as a time-integral of 
the probability to be in the barrier, which in turn decomposed into a 
probability density at each position and time:
\begin{eqnarray}
\tau & \equiv & \int_{-\infty}^{\infty} dt P_{\rm bar}(t) \nonumber \\
& = & \int_{-\infty}^{\infty}dt\int_{0}^{d}dx |\Psi(x,t)|^{2} \; .
\end{eqnarray}
By generalizing this to the case of postselected subensembles (i.e., 
calculating the weak values of the projector $\delta(\hat{X}-x)$ for
various positions $x$), it proved possible to derive a ``conditional 
probability distribution'' for a particle to be at position $x$, {\it 
given} that it was prepared in a state $\ket{i}$ (incident on the 
barrier from the left, in a given wave packet) and detected in a final 
state $\ket{f}$ (transmitted to the far side of the barrier).  The 
time $\tau$ turned out to be in general complex, but its real part -- 
that part which describes the position shift of a pointer coupled to 
the particle's presence in the barrier region -- is of the same order 
of magnitude as the group delay, and exhibits the same 
``superluminal'' features.  A plot of the evolving 
conditional-probability distribution is shown in Figure 3.

One of the striking things about this figure is that the particle 
appears to spend essentially no ``time'' (in the sense of the real 
part of a weak value) near the centre of the barrier.  A reflected 
particle only spends time within an exponential decay length of the 
input facet; while a transmitted particle spends roughly equal amounts
of time near the entrance and near the exit (as one might have 
surmised from the symmetry of the experimental arrangement, or of the 
formula for weak values).  Figure 4 presents a {\it gedankenexperiment}
to elucidate the physical meaning of these curves.  Consider a proton
constrained to tunnel in one direction.  It tunnels
along a series of holes in parallel conducting sheets, which serve
to break the tunnel barrier up into a sequence of electrically-shielded
regions.  As described in the context of the three-box problem above,
one way to measure the weak value of a ``probability'' (or of its time-integral,
a dwell time) is to study the momentum shift of a test charge which
interacts with the particle in question.  Here we imagine an electron, 
initially at rest, between each pair of conducting plates.  We measure
the final momentum of each electron after the passage (reflection or
transmission) of the proton, sorting according to whether the proton was
transmitted or reflected.  On each event, by definition of a weak
measurement, the electrons' momenta are far too uncertain to draw any
conclusions (or else the presence of the electrons would so perturb the
motion of the proton that there would be no sense in discussing it as
a tunneling problem; see \citep{Steinberg=JKPS}).  After averaging over the momenta
found for numerous transmitted protons, however, one would find the
symmetric distribution indicated in the figures.  

In keeping with our intuitions, but not with the standard (time-asymmetric)
recipe for dealing with quantum evolution and measurement, we see that
in addition to concluding from the initial condition (a particle approaching 
the barrier from the left) that the wave packet penetrates roughly one
exponential decay length into the left side of the barrier, one may conclude
from the final condition (a particle exiting the barrier on the right, for
instance) that it had penetrated one decay length into the right side of
the barrier as well.  Weak measurements allow us to discuss the behaviour
of ``to-be-transmitted'' particles and ``to-be-reflected'' particles
separately, and observe that even when described by the same initial
wave function, they may have different physical effects on weakly coupled
environments.

It turns out that one of the popular approaches to tunneling times, the
Larmor time \citep{Buttiker=1983}, is in essence nothing but a weak value.
This time has two different components, whose individual physical meanings
were obscure, however, until reinterpreted in the light of this new
formalism.  It is now clear that they correspond to the real and imaginary
parts of the weak measurement, and that the former corresponds to the
pointer shift (the measurement result as extrapolated from the classical
limit), while the latter indicates a necessary back-action of the particle
due to the measurement, which can be made arbitrarily small by using a
sufficiently weak measurement.  

One question raised by the evolving conditional probability distributions
plotted above is whether, in the superluminal-tunneling regime, the particle
really does move from a wave packet on the left of the barrier to one on
the right in a time shorter than $d/c$, without spending significant time
in the centre of the barrier.  While we all know that a cause cannot have
any measurable effect at a spacelike separated point, is it perhaps possible
for a single particle to have an effect at {\it two} points spacelike
separated from one another (but not from the source of the particle)%
\citep{Steinberg=1998FP}?
Clearly, it suffices for two people on opposite sides of a radio transmitted
to listen to the same broadcast, for a cause to have two spacelike-separated
effects.  But is a single quantum particle truly as nonlocal as this radio
wave?  We all know that if a strong measurement is made of the position
of a photon, it can no longer be found in a different position.  But since
repeated weak measurements can be made on the same wave function, and are
not modified by the action of other weak measurements made at the same
time, I was led to suspect that it should be possible to weakly measure
the probability of a tunneling particle passing through a region of 
spacetime which contains the bulk of the incident wave packet, as well
as the probability of the same particle passing through the (spacelike
separated) region which contains the bulk of the transmitted wave packet.
If conditioned on eventual transmission of the tunneling particle, both
of these would be close to unity -- on average, each individual particle
would have had an effect on two spacelike-separated detectors.  Figure  5
shows a space-time diagram for the experiment under consideration%
\citep{Steinberg=1998SLM,Steinberg=AnnPhys}.   An
energy-filter is added after transmission, to ``erase''%
\citep{Scully=1991,Kwiat=1992} any information
about the time of arrival of the transmitted peak; without this filter,
the possibility of a {\it strong} measurement of the time of arrival
of the particle would preclude any possibility that it had come from the
initial peak.  

We have been setting up an experiment\citep{Steinberg=AnnPhys,%
Steinberg=Garda}
to observe laser-cooled
atoms tunneling through an optical barrier, wherein probes interacting with
atoms at various positions and various points in time should allow us to
study the weak-measurement predictions.  In parallel, we have been
thinking about the theoretical approach necessary to determine whether
each single particle had actually affected two measurement apparatuses
at spacelike separation, or whether despite this appearance on average,
each particle could be thought of as being at only one device at a time.
If the wave function is not merely a measure of our ignorance, but in
some deeper sense ``real,'' then one ought perhaps not to be surprised
by a particle having a (weak) effect in two places at the same time, so
long as no ``collapse'' occurs.  Nevertheless, I believe that most physicists
still have an underlying intuition about the indivisibility of particles
which would lead them to predict such effects could not occur.  Amusingly,
when I have tried to explain our proposed experiments, most of the 
physicists {\it I} know, who are willing to discuss such things, had
the opposite reaction: of course a particle can be in two places at the
same time, and of course both pointers may shift simultaneously!

Our initial proposal was to build on the following idea.  Consider
pointers $P1$ and $P2$ at spacelike separated positions.  We would
like to demonstrate that even though each picks up only a small shift
on a single event, it is possible to show that individual particles
interacted with {\it both} pointers.  Let us therefore assume the
opposite, the corpuscular hypothesis that on a given event, either $P1$
or $P2$ was affected, but not both.  Nevertheless, weak measurements
will show that both $P1$ and $P2$ are shifted on average by an amount
roughly equal to unity (a measurement that the particle was almost
certainly in a given region).  This must imply that on some occasions,
$P2$ is unshifted, while on other occasions, it is shifted by an amount
greater than unity; and the same for $P1$.  Due to the anticorrelation of these
shifts, we expect the distribution of the {\it difference} $P1-P2$ to
develop a larger uncertainty.  If the uncertainty of $P1-P2$ did not
grow, we would conclude that the shifts of $P1$ and $P2$ were not
anticorrelated, and that each individual particle must really have
interacted with both.

While some work has been started on higher-moment weak values%
\citep{Iannaccone=pre}, this field is far from mature.  We decided that
a simple approach would be to use the same measuring device at $P1$
and $P2$, but with equal and opposite signs.  For instance, using
the Larmor-clock approach, a magnetic field along $+z$ at region $P1$
could couple to the electron's spin so long as the particle was in
that region, while a magnetic field along $-z$ at region $P2$ could
couple to the same spin with the opposite sign.  The rotation of the spin
in the $x-y$ plane would automatically record the difference between
$P1$ and $P2$.  It is straightforward to show, in the limit of very
weak measurements and narrow-band energy filters, that the effects
of the two magnetic fields should cancel perfectly.  All the transmitted
particles should have their spin unaffected, implying that they were
affected equally by the two interaction regions.  This would, I thought,
support the hypothesis that quantum particles can truly be in two places
(and have measurable effects there) at the same time.

More recently, consideration of the three-box problem described above
led me to carry out the same calculation in that situation.  Spin
rotations of opposite sign in arms $A$ and $B$ would also cancel out,
implying that the particle was really in {\it both} $A$ and $B$
simultaneously.  Yet we saw earlier that the joint probability for being
in $A$ and $B$ was in fact zero.  One can go through the same
argument in the tunneling case.  Even though the conditional probability
distribution does fill both regions $P1$ and $P2$, the product of
projection operators onto two spacelike separated regions automatically
vanishes (in the Heisenberg picture), because these regions constitute
orthogonal subspaces of Hilbert space.  It now seems that even in the
case of superluminal tunneling, a true weak measurement of the joint
probability of being in two places at once is always guaranteed to yield
zero.  Thus even though $\expect{P1}_{\rm wk} =
\expect{P2}_{\rm wk}$ and $\expect{P1-P2}_{\rm wk} = 0$, one can show
\begin{eqnarray}
\expect{(P1-P2)^2}_{\rm wk} & = & \nonumber \\
& = & \expect{P1^2}_{\rm wk} + \expect{P2^2}_{\rm wk}
- \expect{P1P2}_{\rm wk} - \expect{P2P1}_{\rm wk} \nonumber \\
& = & = 1 + 1 + 0 + 0 = 2 \; .
\end{eqnarray}
If one treats this as the definition of the uncertainty in a weak
value, one certainly finds anticorrelations: $P1$ and $P2$ only shift
by unity at the expense of their difference growing uncertain
by $\sqrt{2}$, just as though they had shifted in an entirely uncorrelated
fashion.  On the other hand, if one simply calculates the final state
of a transmitted spin which was subject to equal and opposite interactions
at $P1$ and $P2$, one finds no increase in the uncertainty of its
orientation.  Further work will be necessary to determine what weak
values can really teach us about nonlocality, and how best to define
the uncertainties and correlations of these probabilities which are
not bounded by the usual classical rules.  Nevertheless, it is apparent
that weak measurements allow us to discuss postselected systems (such as
tunneling particles) in a much more powerful way than was possible
in the more conventional language of evolving and collapsing wave
functions.  In the meantime, we continue to build our laser-cooling
experiment, to verify these predictions, and to study generalisations
which occur when ``real'' measurements (i.e., decoherence or dissipation)
are introduced, and when the ``weakness'' of an interaction is varied.

\section{Quantum information and postselection}

In the burgeoning field of quantum information\citep{Nielsen=2000},
it is well known that photons are excellent carriers of quantum 
information, easily produced, manipulated, and detected, and 
relatively immune to ``decoherence'' and undesired interactions with 
the surrounding environment.  This has led to their widespread 
application in quantum communications \citep{Bennett=1984,Brendel=1999%
,Buttler=2000}.
Unfortunately, the superposition principle of linear optics implies 
that different photons behave independently of one another-- without 
some nonlinearity, it is impossible for one photon to influence the 
evolution of another photon, and this has long made it seem that 
optics would be an unsuitable platform for designing a quantum {\it 
computer}.  Even certain straightforward projective measurements, such 
as the determination of which of the four Bell states\footnote{the 
maximally-entangled polarisation states of two particles:
$\ket{HV}\pm\ket{VH}$ and 
$\ket{HH}\pm\ket{VV}$ in the case of photons, or equivalently,
$\ket{J=0,m=0}$, 
$\ket{J=1,m=0}$, and $\ket{J=1,m=1}\pm\ket{J=1,m=-1}$ for a pair of 
spin-$1/2$ particles.} a photon pair is in, prove to be intractable 
without significantly stronger nonlinearities than exist in 
practice\citep{Mattle=1996,Bouwmeester=1997,Calsamiglia=2001}.  Much work has
focussed on developing exotic systems such as cavity-QED experiments 
\citep{Nogues=1999,Turchette=1995} in which enhanced nonlinearities allow for the 
design of effective quantum logic gates, while most of 
quantum-computation research has instead focussed on using atoms, 
ions, or solids to store and manipulate 
``qubits''\citep{Cirac=1995,Monroe=1995,Kane=1998}.  Recently, it was noted 
that detection {\it itself} is a nonlinear process, and that 
appropriately chosen postselection may be used to ``mimic'' the kinds 
of optical nonlinearity one would desire for the construction of an 
optical quantum logic gate\citep{KLM,Pittman=2001}.  In parallel, work has 
continued on searches for systems in which true optical nonlinearities 
might be enhanced by factors on the order of $10^{9}$ or $10^{10}$, as 
would be necessary for the construction of fundamental logic 
gates\citep{Franson=1997,Harris=1999,Kash=1999}.

We recently showed that it is possible to use quantum interference 
between photon pairs to effectively enhance nonlinearities by a 
similar order of magnitude.  Using a crystal of BBO, beta-barium 
borate, it is possible to frequency-double a beam of light, 
converting two photons at $\omega$ into one photon at $2\omega$ with 
some small (${\cal O}(10^{-10})$) probability, or alternatively to 
``down-convert'' a photon at $2\omega$ into a pair of photons around 
$\omega$, with equally low probability\citep{Steinberg=1996AMO}.  These effects 
are extremely 
common and extremely important in modern nonlinear optics, but rely on 
high-intensity beams to generate significant effects; two individual 
photons entering such a crystal would have a negligible interaction.
For this reason, one experiment which purported to perform ``100\%
efficient'' quantum teleportation by using a nonlinear interaction to 
carry out the necessary Bell-state determination actually needed to 
replace one of the incident photons with a beam containing billions 
of identical copies\citep{Kim=2001}.  By contrast, we discovered 
that adding an additional pump beam (with billions of photons) to the 
system leads to a quantum-interference effect which can enhance the 
interaction between two single-photon-level beams by many orders of 
magnitude.  In \citep{Resch=2001PRL}, we show that this can lead to 
$>50\%$-efficient frequency-doubling of photon pairs.  This effect is closely 
related to earlier work on quantum suppression of parametric 
down-conversion by Anton Zeilinger's group \citep{Herzog=1994}.

The basic scheme is shown in Figure 6.  Two beams at $\omega$, each containing 
less than 1 photon on average, enter a nonlinear crystal; these beams 
are conventionally known as ``signal'' and ``idler.''  Simultaneously, 
a strong pump beam at $2\omega$ pumps the crystal in a mode which 
couples to signal and idler via the interaction Hamiltonian
\begin{equation}
{\cal H} = g a^{\dagger}_{p}a_{s}a_{i} + {\rm h.c.} \; .
\end{equation}
This can convert a single pump photon into a signal-idler pair, or 
vice versa, albeit with vanishingly small efficiency.  The three input 
beams are in coherent states, and thus the initial state of the system 
may be written $\ket{\Psi} = 
\ket{\alpha_{p}}_{p}\ket{\alpha_{s}}_{s}\ket{\alpha_{i}}_{i}$.  For 
weak inputs $|\alpha_{s}|,|\alpha_{i}| \ll 1$, but a strong classical 
pump ($|\alpha_{p}|^{2} \sim 10^{10}$), the interaction can be 
controlled such that to lowest order, all photon pairs are removed 
from the signal and idler beams (i.e., they are up-converted into the 
pump mode, although this effect is too weak to be directly observed).
This occurs due to destructive interference between the amplitude for 
a photon pair to be present in $s$ and $i$, and the amplitude for a 
pump photon to down-convert into the same modes\citep{Resch=2002JMO}.  Importantly, this 
interference effect depends on the relative {\it phase} of the three 
beams, which means that it cannot work if any of the beams has a 
well-defined photon number, since the optical phase and the photon 
number are incompatible observables (roughly speaking -- on the same 
order of roughness as the time-energy uncertainty principle -- 
$\Delta n \Delta\phi \geq 1/2$).

This upconversion effect can be thought of as a highly efficient 
switch-- if a photon happens to be present in the signal mode, than 
no photon in the idler mode can be transmitted; and vice versa.  
Unfortunately, this is only true if it is fundamentally {\it unknown} 
whether the signal mode possessed a photon or not.  By observing the 
absence of coincidence counts after the device, we may conclude that 
any photon pairs which had been present disappeared\ldots but on no 
individual occasion did we know a photon pair actually existed!

We extended this work to a geometry more closely related to one of 
the standard logic gates of quantum information theory, the 
controlled-phase gate \citep{Nielsen=2000,Resch=2002cphi}.  Still relying on 
interference between incoming photon pairs and the down-conversion 
process, we altered the relative phase so that the {\it probability} 
of a photon pair emerging was not significantly altered, but its 
quantum phase would be shifted relative to that of the vacuum or a 
single photon in either beam alone.  To measure this, we built the 
homodyne setup in Figure 7.  This can be thought of as a simple 
Mach-Zehnder interferometer for a signal photon (really a signal beam 
with an average photon number per pulse much less than 1).  Into one 
arm of the interferometer, our pumped crystal is inserted.  At the 
same time, a ``control'' beam is sent through the crystal's idler 
mode.  If a control photon is present, then a phase shift is 
impressed on any passing signal photon; this is observed as a shift 
in the Mach-Zehnder interference pattern (see Fig. 8).  We were able 
to observe shifts as large as $\pm 180^{\circ}$, or very small phase 
shifts with little effect on the probability itself, depending on the 
strength of the pump beam relative to that of the signal and control 
beams.  Once more, however, to operate this gate, we had to operate 
in a condition of ignorance.  We send in beams which may or may not 
have photons, but when we observe a ``control'' photon leaving the 
crystal, we find the desired effect on the signal.  Can we conclude 
that this postselection determined that there had been a control 
photon there all along, and that the logic gate performed the correct 
operation for an input of logical `1'?   To understand the operation 
of the gate -- the phase shift imprinted on the signal beam -- it is 
necessary to take into account both the state preparation (the 
well-defined phase differences between the beams) and the 
postselection (the presence of a control photon). 

In a manner somewhat reminiscent of the KLM scheme \citep{KLM}, this 
requires a fundamental change 
in the way one thinks about logic operations, with inputs being 
determined not by preparing the appropriate state, but by 
postselecting the desired value of the input \citep{Resch=2002qsp}.  
So far, it remains 
unclear how widely such effects could be applied in quantum 
information; we do not presently know of a way to incorporate them 
into the standard paradigm of quantum computing.  On the other hand, 
we have shown\citep{Resch=Solvay} that despite its eccentricity and 
potential pitfalls, this ``conditional-phase switch'' can indeed be 
used to implement the Bell measurements which were previously 
impossible for individual photon pairs, provided only that the photon 
pairs are produced in the appropriate superposition with vacuum.  For 
subtle but important reasons, this means our technique {\it cannot} 
be used for unconditional quantum teleporation; but it can be used to 
improve earlier experiments on subjects such as quantum dense 
coding\citep{Mattle=1996}. 

 While it is not 
possible to have a well-defined phase {\it and} a well-defined photon 
number in a quantum state, it is possible to prepare one and 
postselect the other: and weak measurements show us that at 
intermediate times, the system possesses some characteristics of both 
the initial and final states.  It seems that weak measurement may be 
precisely the formalism needed for describing such enhanced 
nonlinearities, and probably a broader range of ``nondeterministic'' 
operations currently being investigated in quantum logic.

\section{Having your cake and eating it too}

There is another example of a possible application for these enhanced 
nonlinearities, and we are presently setting up an experiment to
demonstrate this.  In 1992, Lucien Hardy proposed an ingenious quantum
paradox which involved intersecting electron and positron 
interferometers, wherein colliding electrons and positrons would 
undergo certain annihilation\citep{Hardy=1992realism,Hardy=1992PRL}.  
Of course, this scheme was  
quickly recognized to be something of a stretch experimentally, and
it was hoped that the experiment could be performed with optical 
interferometers instead.  Unfortunately, as mentioned several times 
already, the interaction between different photons is so weak in 
practical systems that the equivalent of an ``annihilation'' event -- 
an upconversion event, for instance -- was exceedingly rare.  A 
mathematically equivalent paradox was eventually  tested optically 
\citep{White=1999}, but no direct demonstration of the original 
conundrum has been possible to date.

Hardy's paradox relies on the concept of ``interaction-free 
measurements'' introduced by Elitzur and Vaidman \citep{Elitzur=1993}.
Briefly, it is possible to set up an interferometer as in Figure 9 to 
transmit all the input light out one port, known as the ``bright'' 
port.  Ideally, no photon should ever be detected at the ``dark'' 
port.  However, any object which blocks one of the paths of the 
interferometer will destroy the interference, and therefore generate
some probability of a photon exiting the dark port.  Clearly, in the 
cases in which a photon is observed at this port, one can conclude 
that (A) it was not blocked by the object; but (B) the object must 
have been in place (since without the object, interference prevents 
any counts from being observed there).  In the original example, this 
made it possible to achieve the surprising feat of confirming that an 
infinitely sensitive bomb was functioning -- without setting it off.  
In later work \citep{Kwiat=1995}, it was shown that this task could be 
accomplished with arbitrarily high efficiency, through ingenious 
modifications to the interferometer.

Although such measurements are popularly referred to as 
``interaction-free,'' in some quantum mechanical sense, they clearly do 
involve an interaction: a ``bomb'' initially in an uncertain position 
may be collapsed into the interferometer arm, through the detection 
of a photon at the dark port.  Such considerations motivated the 
extension of the problem to two overlapping 
interaction-free-measurement (IFM) interferometers, shown in Figure 10, each 
of which can be thought of as measuring whether or not the {\it other} 
interferometer's particle is in the ``in'' path.  The reasoning now is 
simple.  If an electron interferometer and a positron interferometer 
overlap at ``in,'' in such a way that the electron and positron are 
certain to annihilate if they meet there,
 then each particle may serve to ``block'' the other
particle if and only if it takes the ``in'' path.  If each 
interferometer is aligned so that all electrons reach $B_{e}$ and all 
positrons $B_{p}$, then these two interferometers are IFMs.
An electron will only be detected at $D_{e}$ 
if the positron was in the way.  Similarly, a positron can only reach 
$D_{p}$ if an electron is in the way.  Naturally, if both the 
electron and the positron are at ``in,'' then they annihilate, and 
cannot be observed.  For this reason, one should never observe an 
electron at $D_{e}$ {\it and} a positron at $D_{p}$ at the same time.

Yet this is not the case.  Quantum mechanically, there is a finite 
probability for both the electron and the positron to reach their dark 
ports.  How do we interpret this?  The conventional answer is that we 
have learned the error of our classical ways.  While the IFM was able 
to tell us whether or not a classical particle was blocking one arm 
of an interferometer, we transgressed by drawing counterfactual 
conclusions about a quantum particle which was not directly 
observed.  Clearly, this is not a very satisfying state of affairs, 
but perhaps it is true that quantum mechanics does not allow us to 
make ``retrodictions'' of the sort we rely on to construct this 
paradox.

Despite the clear contradiction with classical reasoning, the astute 
reader may recall that at least in the case of weak measurements, 
classical intuition often works surprisingly well, albeit at the 
expense of certain other intuitions, such as the positive-definiteness
of probabilities.  Indeed, it was recently pointed out \citep{Aharonov=2002}
that weak measurements can ``resolve'' the 
paradox raised by Hardy.  How is this?  Consider weak measurements of 
the probabilities for the various particles to be in the various arms 
of the interferometer, and of the corresponding joint probabilities.  
From where does the apparent paradox arise?  If we post-select on 
cases where both photons reach the dark port, we want to conclude that 
the probability of the electron having followed the ``in'' path, 
$P(e^{-}{\rm in}) = 1$; and also that $P(e^{+}{\rm in}) = 1$.  So far 
so good, except that we also believe that $P(e^{-}$  in {\it 
and} $e^{+}$ in$)$ must $= 0$, since both particles would have 
annihilated had they met along the ``in'' path.  Of course, we have 
already seen a similar situation in Table 1.  Just because A and B 
both happen with certainty (in a weak-measurement sense) does {\it 
not} imply that A and B ever happen simultaneously.  Aharonov {\it 
et al.} calculate that the above probabilities do in fact hold, and 
that to satisfy the various sum rules, the probability of one 
particle being ``in'' and the other being ``out'' is $100\%$, and
that the probability of {\it both} particles being ``out'' is $-100\%$.
In this sense, there is no more paradox.  All the paths can be 
measured simultaneously and in arbitrary combinations, so long as the 
measurements are all weak.  And given this proviso, all our 
expectations from intuitive analysis of the IFMs should prove to be 
correct.  The price we need to pay for this resolution is to accept 
that, at least in situations of postselection, certain probabilities 
may turn out to be negative.

Although we still do not know how to turn our ``switch'' into a 
quantum computer, recall that it allows us to cause photon pairs to 
upconvert with nearly unit efficiency.  This is the analog of the 
$e^{+}e^{-}$ annihilation in Hardy's original formulation, and we can 
now hope to observe his paradox directly, using a coherently-driven 
nonlinear crystal as the interaction region for ``annihilating'' our 
photon pairs.  Now this switch, which had the disturbing property of 
working only in a ``nondeterministic,'' after-the-fact manner, 
becomes the ideal tool for studying the difficult situations one gets 
into when trying to make retrodictions about quantum-mechanical systems.

\section{Conclusion}

In this rapid tour of a variety of recent (and future) experiments 
and theoretical investigations, I have tried to focus some attention 
on the new trend towards attempting to talk about {\it history} in 
quantum mechanics, and in particular to talk about the history of 
specific subensembles defined by both state preparation and 
postselection.  The formalism of weak measurements addresses such 
problems in a very natural fashion, but yields all manner of 
counterintuitive predictions.  At the same time, it has an unshakable 
connection to real measurements which could be (and often are) 
performed in the laboratory; I describe certain experiments now in progress 
which should further demonstrate the fruitfulness of this formalism.  
The relationship between weak measurements and generalized 
probability theories appears to be particularly strong, but more work 
remains to be done to elucidate the meaning of these exotic (negative, 
or even complex) quantities which obey many of the axioms of 
probability theory.  In particular, weak measurements provide one with 
a little more leeway than orthodox quantum mechanics when it comes to 
describing what the state of a system ``really was'' between 
preparation and detection, but in so doing, raises a variety of 
difficult questions, especially relating to the reality of the wave 
function, and the nonlocality of individual quantum particles.  It is 
interesting to note that a variety of experiments, ranging from new 
concepts for quantum computation to cavity-QED studies of open-system 
quantum dynamics, have recently provoked increased interest in the 
mathematical description of post-selected subensembles.  Perhaps the 
time is finally right for mainstream quantum physicists to attack 
these problems, and in the process develop a better understanding of 
the nature of space, time, and measurement in quantum mechanics.

\section{Acknowledgments}

I would  like to 
acknowledge my coworkers at Toronto -- Kevin Resch, Jeff Lundeen, 
Stefan Myrskog, Jalani Fox, Ana Jofre, Chris Ellenor, Masoud 
Mohseni, and Mirco Siercke -- both for their efforts in the lab 
and for their ideas, many of which have found their way into this
chapter. 
I would also like to thank Ray Chiao and Paul Kwiat for their collaboration 
and useful discussions over many years.  
 Finally, for thought-provoking conversations about weak measurements
 I would like to thank Howard 
Wiseman, Jeff Tollaksen, Sandu Popescu, Lev Vaidman, Avshalom Elitzur,
Gonzalo Muga, Markus B\"uttiker, 
and Jeeva Anandan, in addition of course to Yakir Aharonov, whose seminars 
first introduced me to the concept.  Some of the work described in this chapter 
was supported by NSERC, by Photonics Research Ontario, by the Canadian 
Foundation for Innovation, and by the US Air Force Office of Scientific
Research under the QuIST programme (F49620-01-1-0468).

\section{Figure Captions}

\begin{enumerate}
\item
A two-slit experiment.
When a particle is fired from a source towards the double slit, we 
can use Schr\"odinger's equation to predict its state as it passes the 
two slits: a symmetric wave function localized equally behind both 
slits.  But when a particle appears at one point on the screen, what
can we conclude about its history?  As we all know, we cannot state 
it went through one slit or the other.  Shall we say it went through
both with equal likelihoods, as determined by the state preparation?  
Or from the location of the spot on the screen, can we construct some 
more accurate wave function?  Can we just use Schr\"odinger's 
equation to propagate the electron backwards in time?  This would 
discard all information about the state preparation, which seems 
extreme.  Yet to discard all information about the future may also be 
unnecessary-- for instance, even the claim of a symmetric 
double-peaked wave function only made sense given the knowledge that 
the particle did make it through the double-slit to eventually reach 
the screen, knowledge only obtained via postselection.\\

\item
a. The quantum 3-box problem.  If a particle is hidden in three
boxes in a superposition $(A+B+C)/\sqrt{3}$, but is subsequently found
to be in the (different but nonorthogonal) superposition 
$(A+B-C)/\sqrt{3}$, what can one say about the state of the particle
while in the box?\\
b.  Experimental schematic for an optical implementation of the 
three-box problem.  Photons are sent into a 3-rail interferometer,
with the three rails playing the roles of boxes A, B, and C.  A $\pi$
phase shift is introduced in rail C, such that detection at the
camera post-selects a superposition $(A+B-C)/\sqrt{3}$.  To weakly
``measure'' the particle in one or another of the boxes, small
transverse displacements are induced in each of the rails, and an
image of the postselected photon distribution is taken to determine
the size of the effects of displacements in each of the boxes.\\

\item
The time-evolution of the ``weak'' conditional probability distribution 
for a particle's position as it tunnels through a barrier.  The heavy
curve shows the real part of this distribution (the magnitude of the
expected measurement result), while the dashed curve shows its
imaginary value (the ``back-action'' due to measurement), and the light
curve shows the distribution for reflected particles (essentially
equal to $|\Psi|^{2}$).  Note that at early and late times, the
weak distribution mimics the full incident or transmitted wave packet,
while at intermediate times it has an exponentially small magnitude
inside the forbidden region.\\

\item
A {\it gedankenexperiment} using distant electrons to measure how much time a
tunneling proton spends in each of several shielded regions of space. While 
the proton is between a given pair
of conducting plates, only the corresponding electron feels a significant force. 
After the tunneling
event, the momentum shift of each electron thus records the amount of time spent by the proton
between the plates in question. The implication of weak-measurement theory is that 
reflected
protons only transfer momentum to electrons near the entrance (a), while transmitted 
protons affect
electrons near both edges of the barrier (b). Electrons in the center only undergo a position 
shift, related to the back-action of the measurement.\\

\item
A gedanexperiment to investigate whether or not a subset of tunneling 
particles may truly prove to have ``been'' in two places at the
same time, due to the superluminal group velocity in tunneling.
The peak of the transmitted gaussian may emerge at a point spacelike
separated fromm the peak of the incident gaussian.  An energy filter
is necessary to ``erase'' any timing information which would preclude
the detected particle from having been present at the incident peak;
once a particle is transmitted through a narrowband filter, 
information about its time of origin is smeared out.\\

\item
The two-photon ``switch'' experiment: quantum interference between 
photon pairs being generated through down-conversion and already
being present in two laser beams can lead to nearly unit-efficiency
upconversion of photon pairs from classical beams.\\

\item
The two-photon switch incorporated into a Mach-Zehnder interferometer
serves to demonstrate a conditional-phase gate, i.e., cross-phase
modulation at the single-photon level.\\

\item
Fringe patterns observed at the output of the Mach-Zehnder when a
trigger photon was detected (black circles; solid line), versus when
no trigger photon was detected (white squares; dashed line).  A
significant phase shift is observed on the signal beam due to the
presence of a single photon in the trigger mode.\\

\item
A Mach-Zehnder interferometer as proposed by Elitzur and Vaidman for
performing ``interaction-free measurements.''  When the path lengths
are balanced, all photons reach the ``bright'' port and none the
``dark'' port.  An absorbing object placed inside the interferometer may cause
photons to reach the dark port, indicating the presence of the object
even though those photons could (in some sense) never have interacted
with the object directly.\\

\item
Two overlapping interaction-free measurement devices (``IFMs,'' in 
the jargon) implement Hardy's Paradox.  One device is an electron
interferometer, and the other a positron interferometer.  They
overlap at W, where it is supposed that an electron and a positron
arriving simultaneously will annihilate with certainty.
If one can truly conclude
from electron detection at $D_{-}$ that the positron was in the
interaction region W, and from positron detection at $D_{+}$ that
the electron was in W, then one should never see coincident 
detections between the two dark detectors, since the particles
would have annihilated at W.  Quantum mechanics shows that this
is not the case.\\

\end{enumerate}

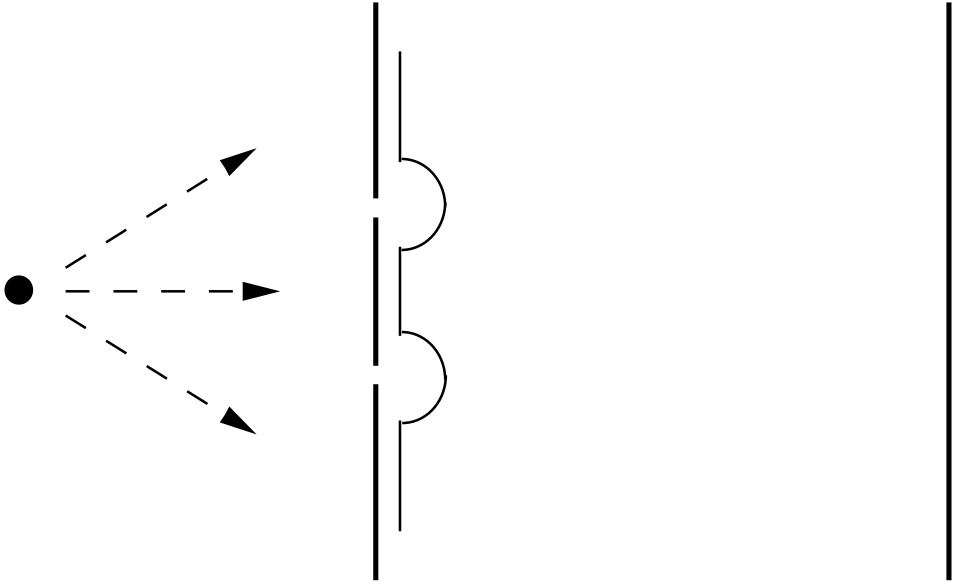
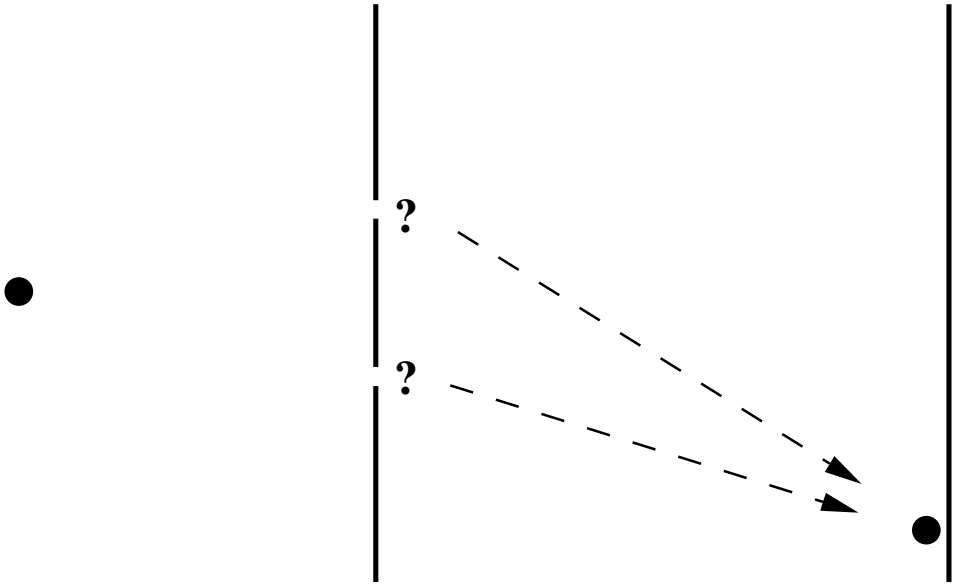
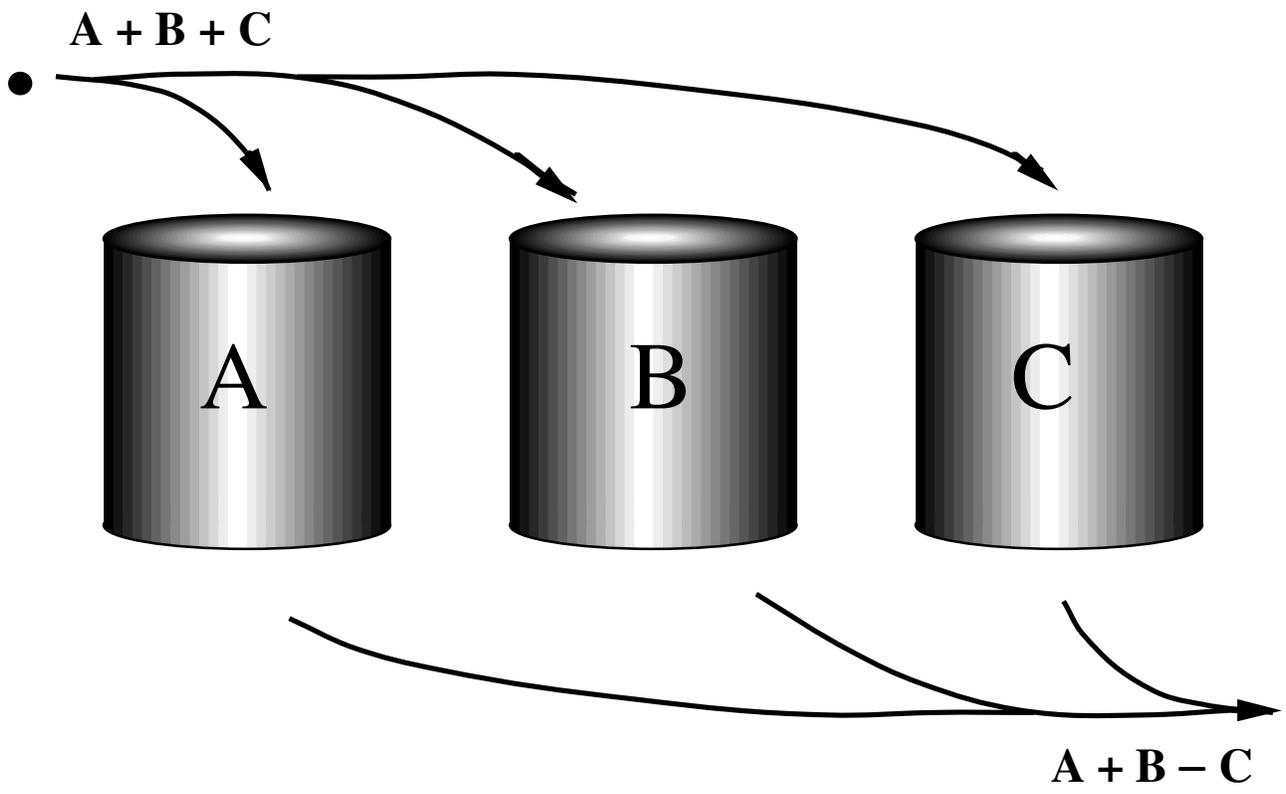
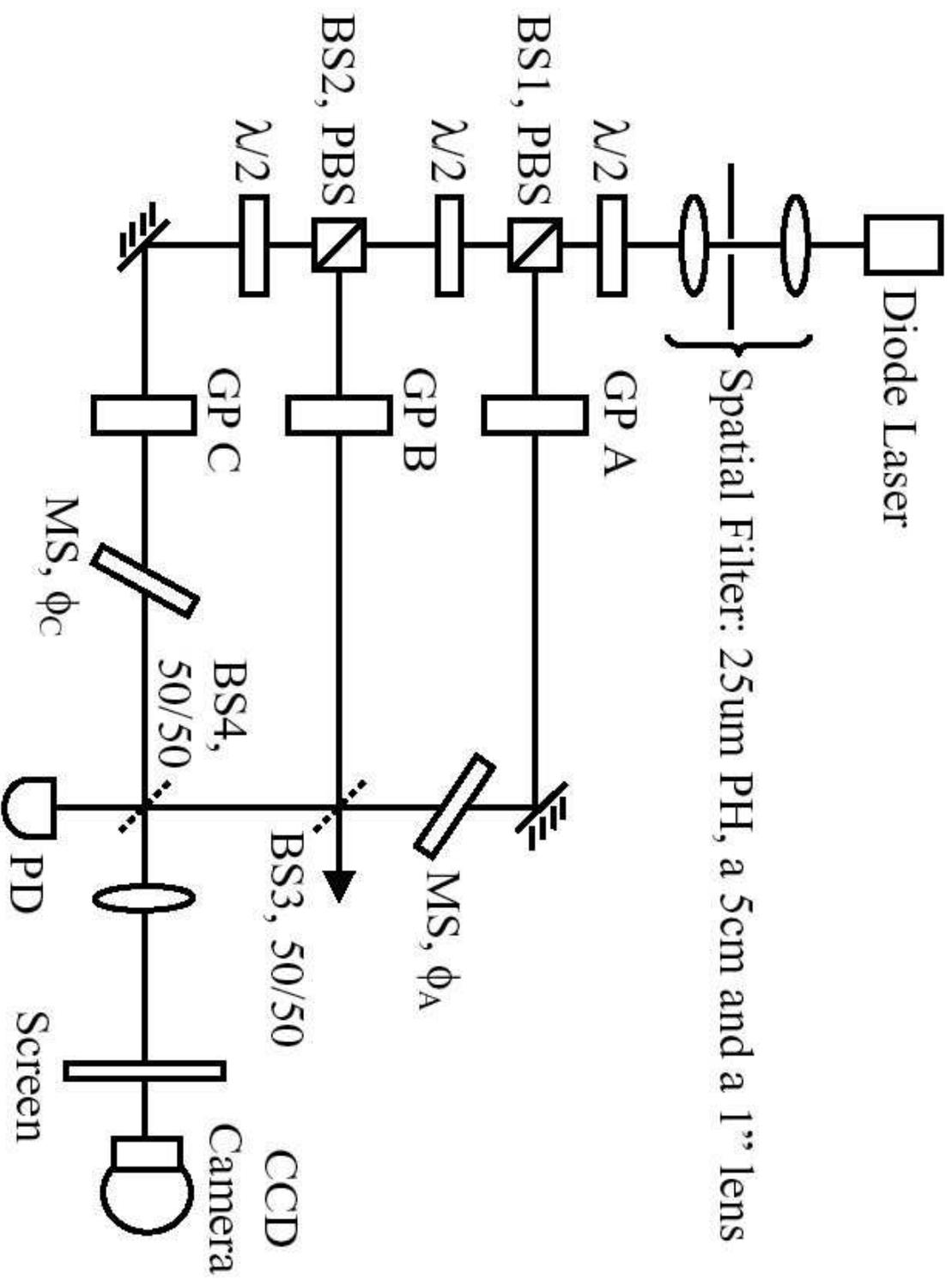
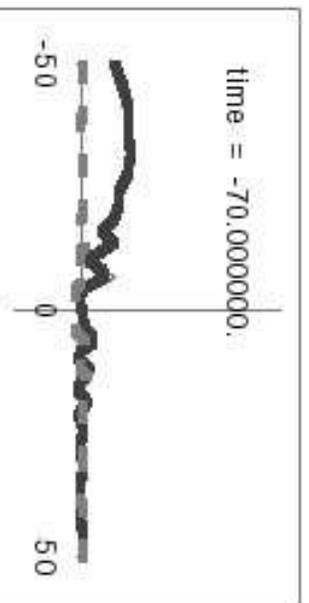
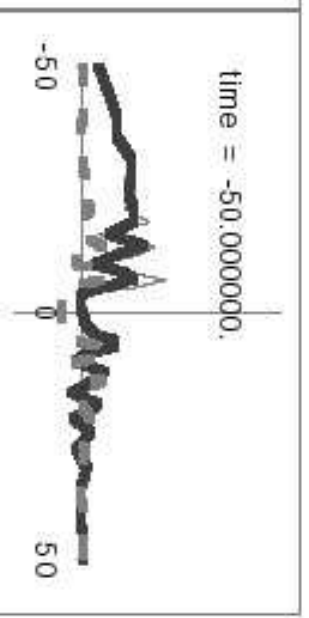
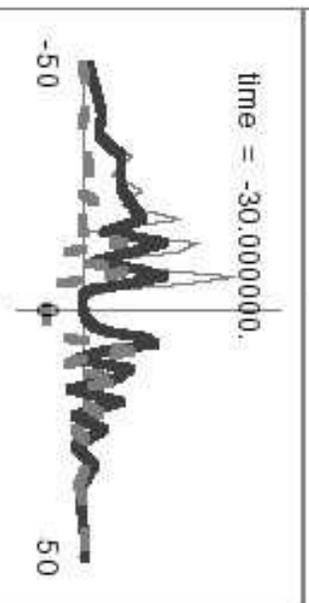
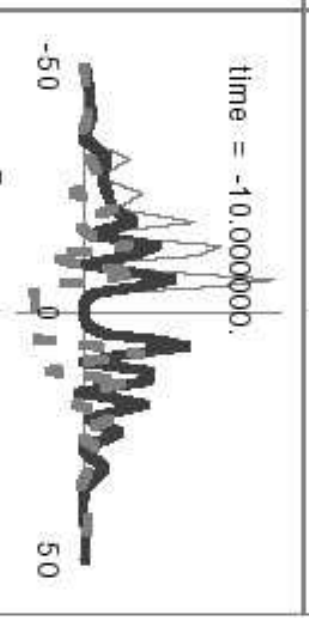
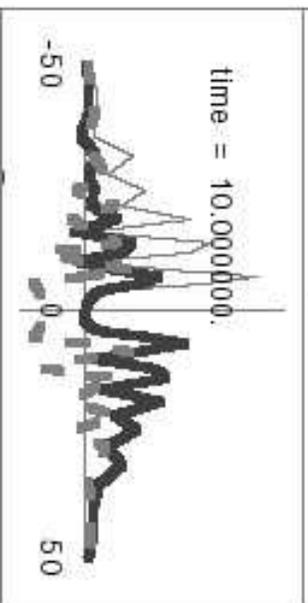
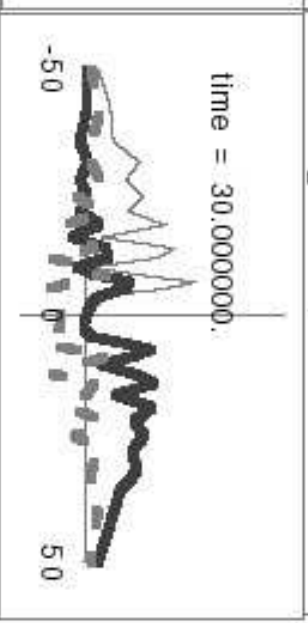
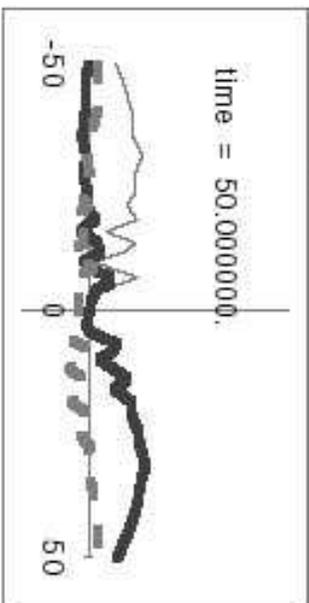
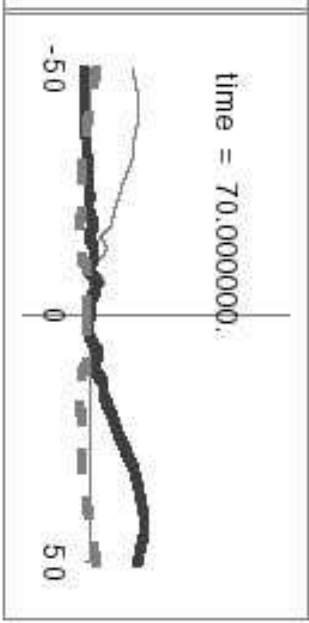
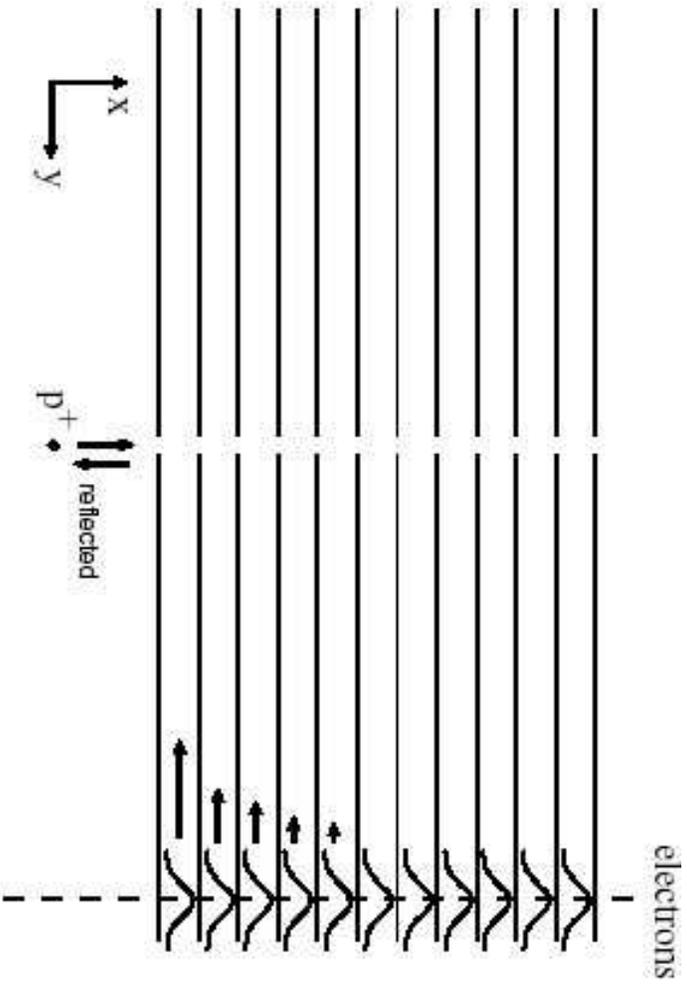
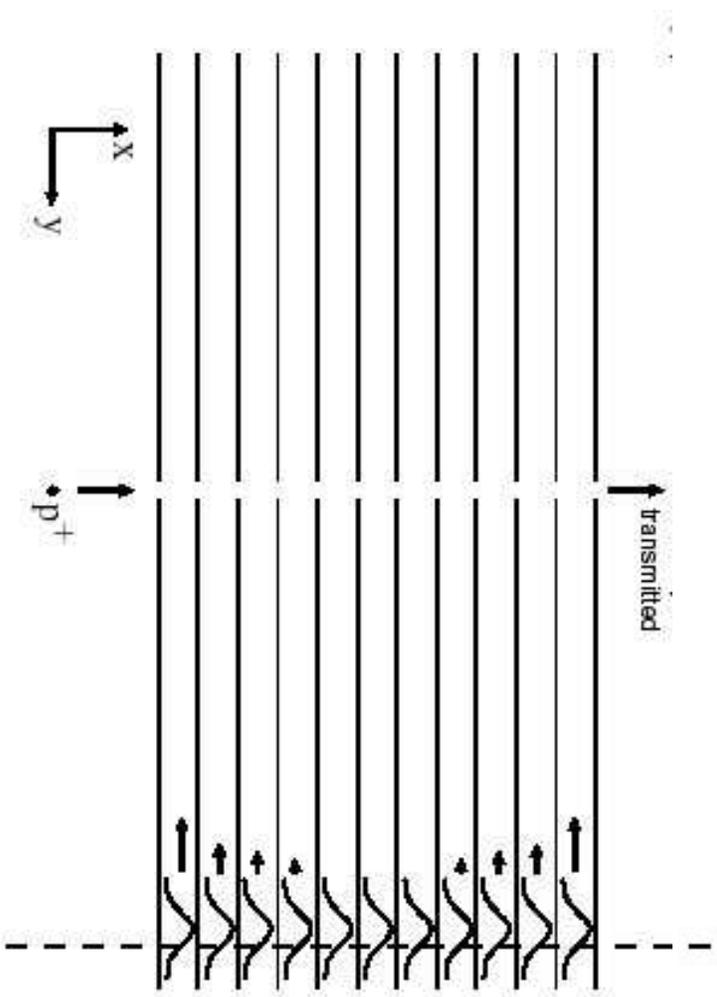
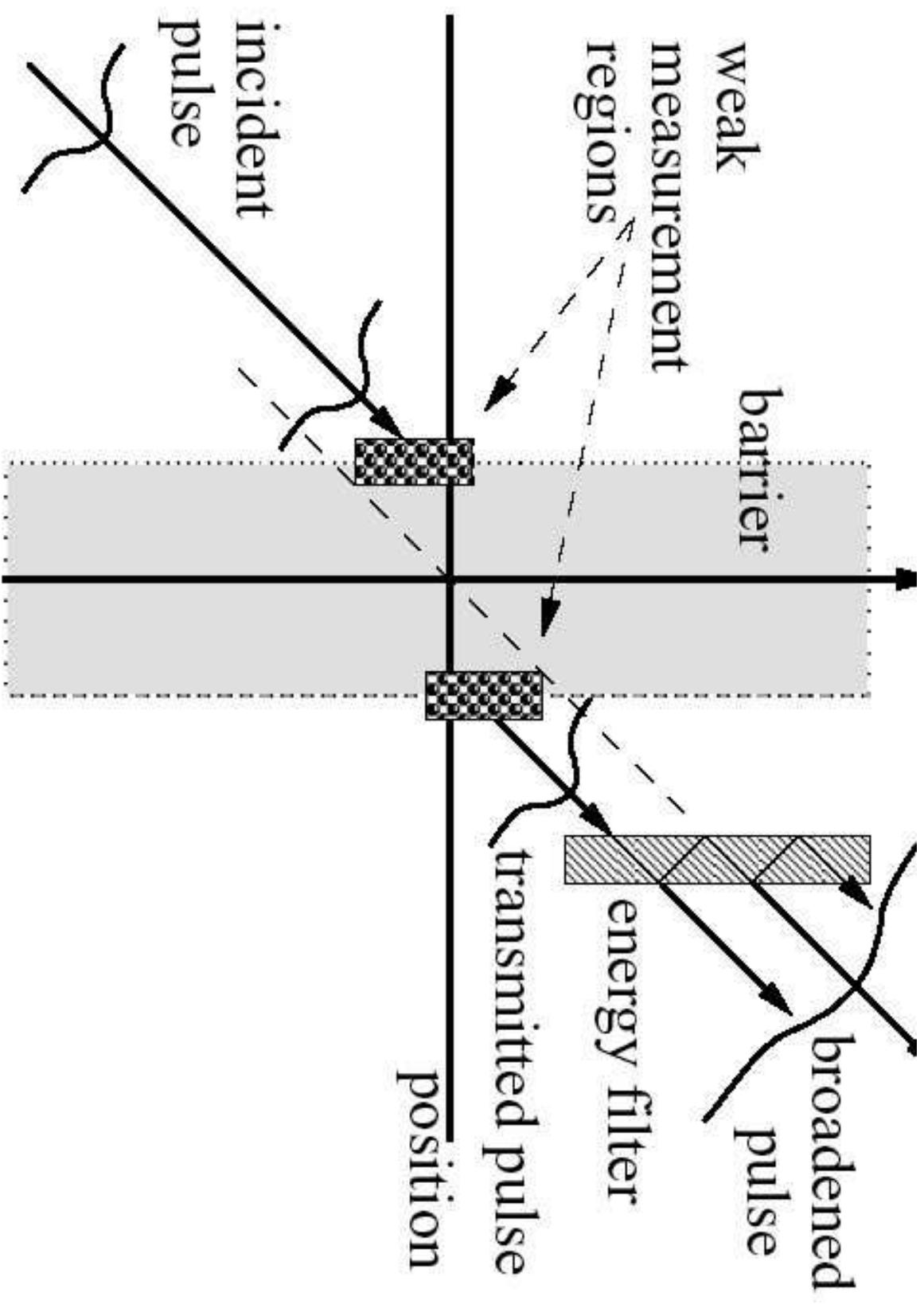
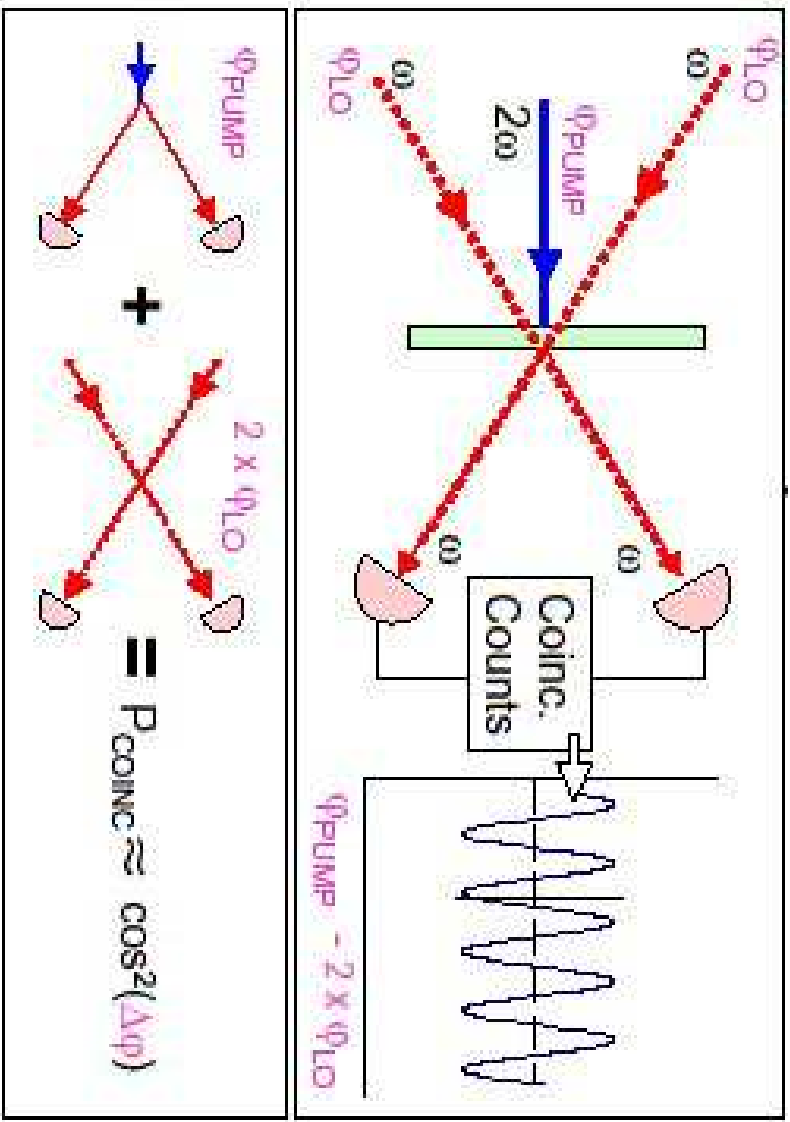
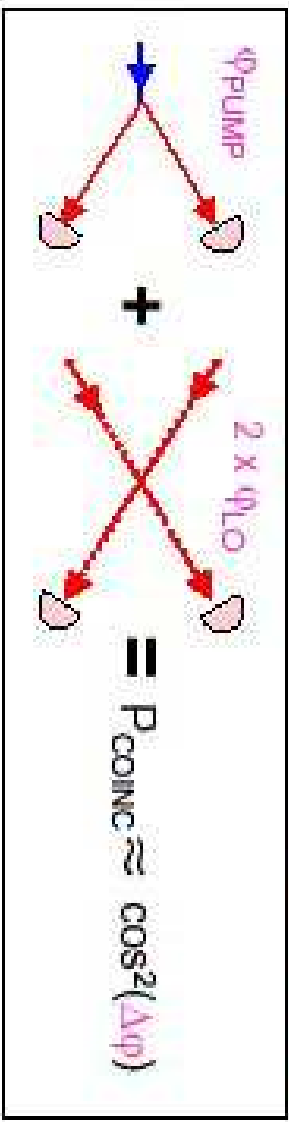
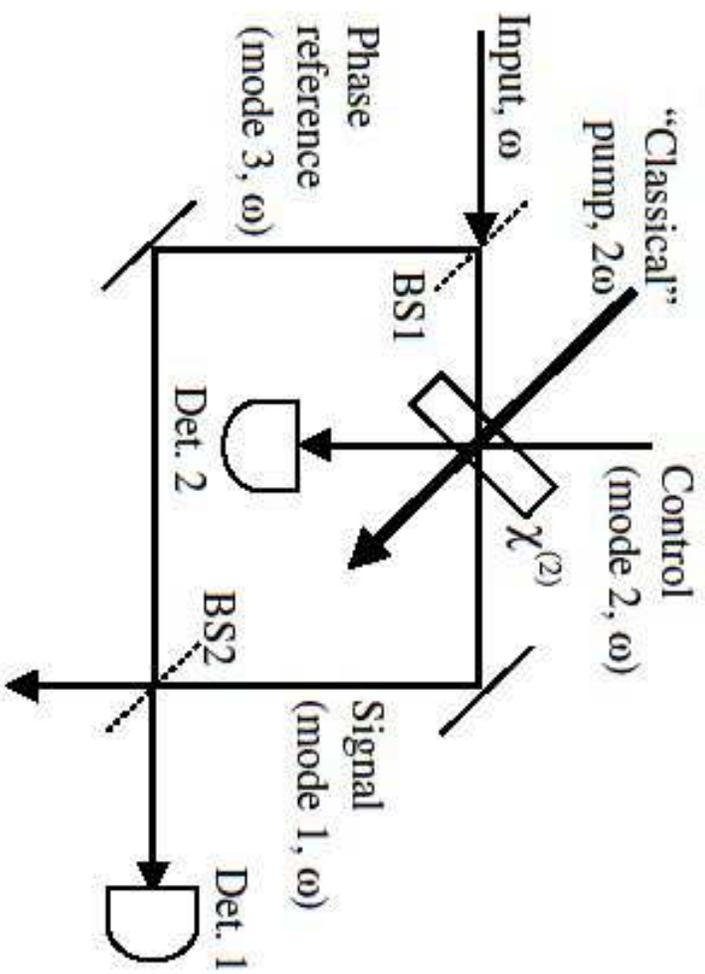
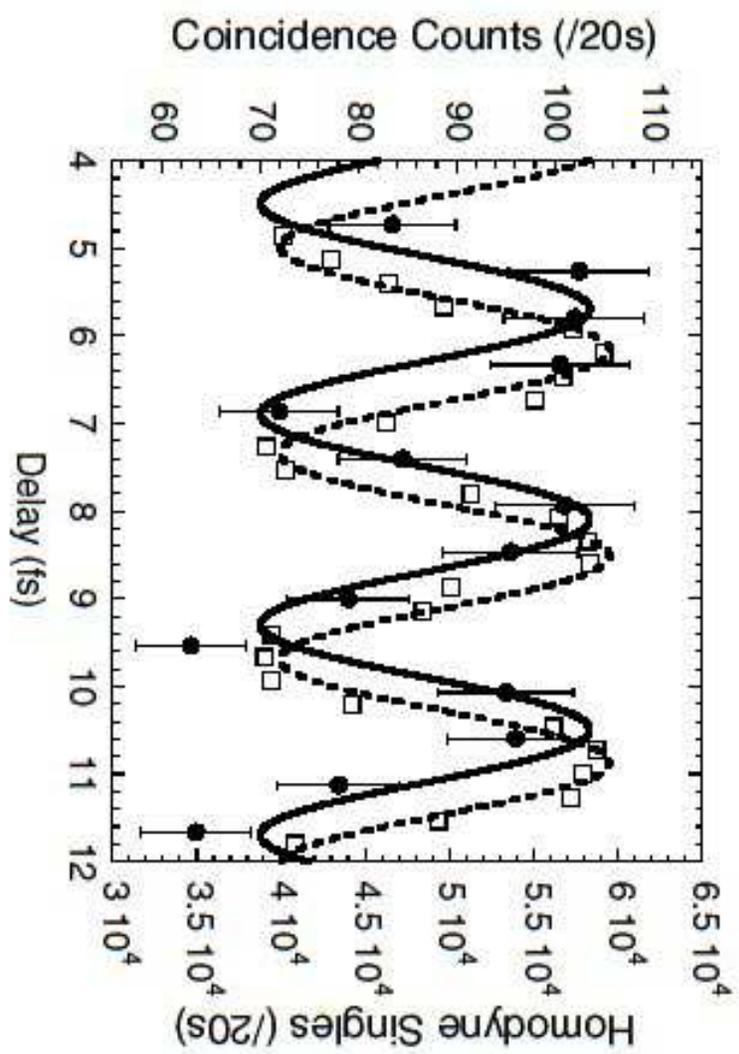
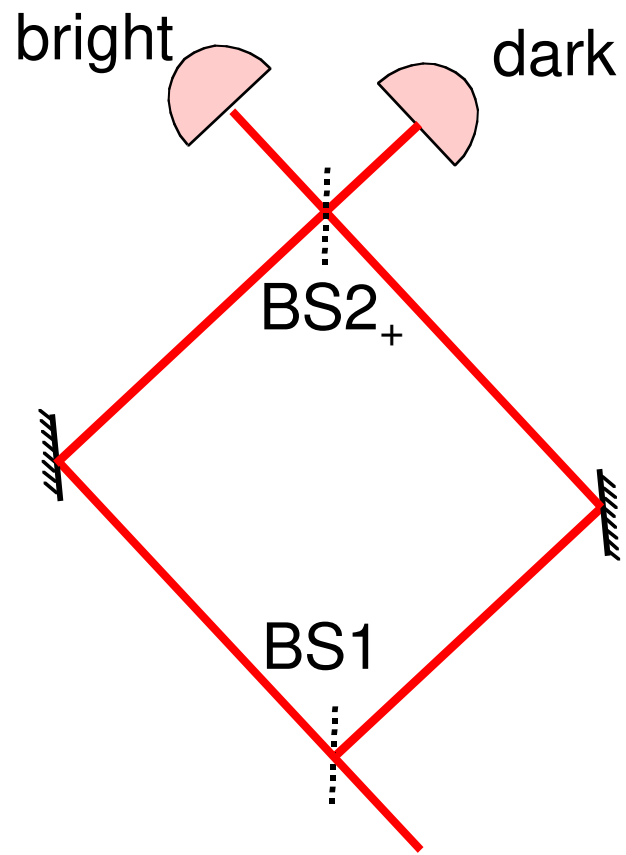
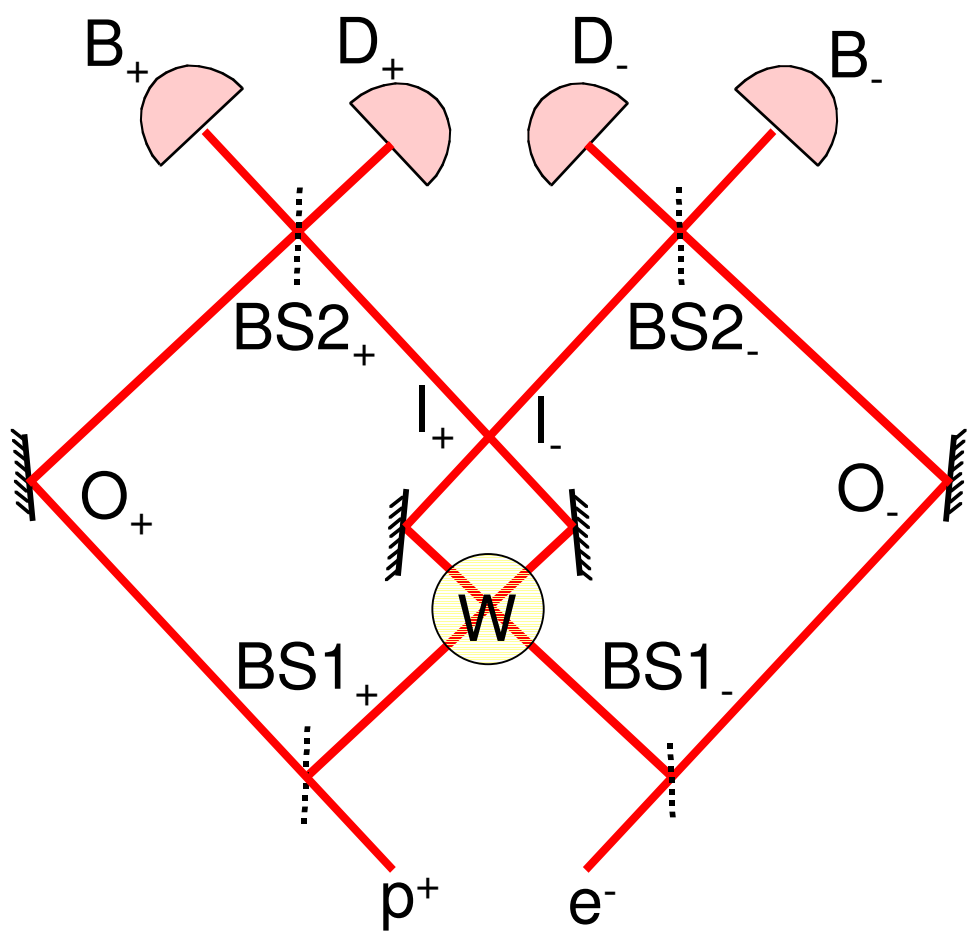
\end{document}